\documentclass[final,5p,times,twocolumn, nopreprintline]{elsarticle}

\journal{Nuclear Instruments and Methods in Physics Research - section A}

\usepackage{amssymb}
\usepackage{amsmath}
\usepackage{lineno}
\usepackage{xcolor}
\usepackage{graphicx}
\usepackage{enumitem}
\usepackage{subcaption}
\usepackage{placeins}
\usepackage{isotope}
\usepackage{comment}
\usepackage{url}
\usepackage[hidelinks]{hyperref}
\newcommand{\nlhref}[1]{\href{#1}{\nolinkurl{#1}}} %automatically create exact url with href
\usepackage{textgreek}

\begin{document}

\begin{frontmatter}

\title{\textbf{A Virtual Frisch-Grid Geometry-Based CZT Gamma Detector for In-Field Radioisotope Identification}}

\author{P.~Maj}
\corref{cor1}
\fnref{label1, label2}
\ead{pmaj@bnl.gov}
\author{G.~Pinaroli\fnref{label1}}
\author{A.~Bolotnikov\fnref{label1}}
\author{G.~A.~Carini\fnref{label1}}
\author{G.~W.~Deptuch\fnref{label1}}
\author{D.~S.~Gorni\fnref{label1}}
\author{S.~Herrmann\fnref{label1}}
\author{D.~Pinelli\fnref{label1}}
\author{J.~Pinz\fnref{label1}}
\author{A.~Verderosa\fnref{label1}}

\affiliation[label1]{organization={Instrumentation Department, Brookhaven National Laboratory}
,
            addressline={PO Box 5000}, 
            city={Upton},
            postcode={11973-5000}, 
            state={NY},
            country={USA}}

\cortext[cor1]{Corresponding author}

\fntext[label2]{This manuscript has been authored by employees of Brookhaven Science Associates, LLC under Contract No. DE-SC0012704 with the U.S. Department of Energy. The publisher by accepting the manuscript for publication acknowledges that the United States Government retains a non-exclusive, paid-up, irrevocable, world-wide license to publish or reproduce the published form of this manuscript, or allow others to do so, for United States Government purposes.}

\begin{abstract}
We present a Virtual Frisch-Grid geometry-based CZT gamma detector developed for identifying different radioisotopes over an energy range from single\,keV up to 2\,MeV, and useful for efficient characterization of CZT crystals. The detector is built with a 3\,\(\times\)\,3 matrix of CZT crystals, each measuring approximately 6\,mm\,\(\times\)\,6\,mm\,\(\times\)\,15\,mm. The charge generated within the sensor’s active volume is read out via an anode connected directly to the AVG3\_Dev integrated circuit. A current signal induced by charge drift is collected on side pads of the crystals, enabling reconstruction of a 3D interaction position. This paper discusses the design, development, and performance of the standalone, mobile detector system, which integrates the AVG3\_Dev readout IC developed at Brookhaven National Laboratory, high-speed FPGA-based with per-channel digital signal processing, and embedded system capabilities. The device is compact, battery-powered, and supports wireless data streaming, making it suitable for field operations towards radioisotope identification.
\end{abstract}


\end{frontmatter}

% \linenumbers

\section{Introduction}
\label{sec:intro}
The development of compact and lightweight gamma-ray detectors is crucial for a wide range of applications, including nuclear nonproliferation, safeguards, environmental monitoring, and security dosimetry. These detector systems integrate high-efficiency, high-energy resolution detectors with ASIC-based readout electronics and signal processing, all within handheld or portable instruments used for isotope identification. Two types of room-temperature semiconductor materials - CdZnTe (CZT) and TlBr - are considered the most promising for such systems \cite{97e69526cbca4e83bb52a5e7ebe8f16d}. While CZT currently dominates this application, TlBr has recently gained significant attention as a material of choice for handheld detectors due to its lower cost and higher detection efficiency~\cite{HITOMI1999160}. Although high-quality CZT crystals are commercially available and are being used in various gamma-ray detector systems, challenges related to handheld instrument requirements - such as low power consumption, rigidity, broad temperature range, and advanced functionalities - continue to drive further development in this field.

The front-end and signal processing electronics, which must accommodate the differing properties of the detector materials across various applications, constitute the most critical and evolving subsystem in these instruments. Using arrays of CZT or TlBr bars, instead of large crystals, offers a cost-effective approach to integrating large-area and volume detectors \cite{Bolotnikov2020b}. The production yield of high-quality CZT and TlBr, especially for larger single crystals, remains a limiting factor. By employing an array of position-sensitive detectors, it is possible to achieve the same or even better performance in terms of effective area and sensitivity with smaller, generally less expensive crystals. Additionally, arrays of smaller crystals can be scaled to cover larger areas and volumes, providing greater flexibility in configuring detectors to meet specific user requirements. Moreover, arrays of individual bar-shaped detectors enable modular design and detector replacement without disrupting the entire system.

In this work, we present the development of the readout system for the position-sensitive virtual Frisch-grid detectors, proposed for a Radioisotope Identification Device (RIID) prototype. The RIID is a lightweight, compact, low-power, handheld system that integrates CZT or TlBr detectors with ASIC-based readout electronics and signal processing. Traditional solutions often face challenges related to integration complexity and high costs. This work aims to address these challenges by introducing a robust, standalone detector system based on the AVG3\_Dev IC and a Frisch-grid architecture. Gamma detectors based on CZT crystals offer unique advantages, including high energy resolution and room-temperature operation \cite{Lordi2013}. However, achieving uniform performance across multiple crystals requires advanced readout and processing systems. Our goal is to address this challenge by presenting a comprehensive system design that integrates cutting-edge hardware and software components, creating a flexible platform for gamma-ray detection and analysis.

The position-sensitive virtual Frisch-grid (VFG) detectors have been proposed to maximize the performance of large-volume bar-shaped CZT crystals with long drift lengths. Achieving high 3D position resolution is critical for enhancing the spectral and imaging performance of these detectors, which operate as time-projection chambers (TPCs). The crystal geometry and detector design offer cost-effective integration of large-area arrays as well as the flexibility to scale both the quantity and size of the crystals, resulting in high sensitivity and better imaging. With 3D position sensitivity, we address the non-uniformity in detector response, presenting a solution to one of the major technological challenges that limit the use of large-thickness and volume CZT detectors in practical applications. The advantages of VFG detectors for spectroscopy and imaging make them highly competitive with comercially available products like e.g. H3D pixelated detectors~\cite{H3Dref}%, \textcolor{red}{add here more} 
. Later in this section, we review the development of VFG detectors in our lab based on CZT, TlBr, and CsPbBr\textsubscript{3} semiconductors, and their potential applications across diverse fields: basic science, medical and industrial imaging, nonproliferation, safeguards, and environmental monitoring.

\begin{figure}
    \centering
    \includegraphics[width=0.7\linewidth]{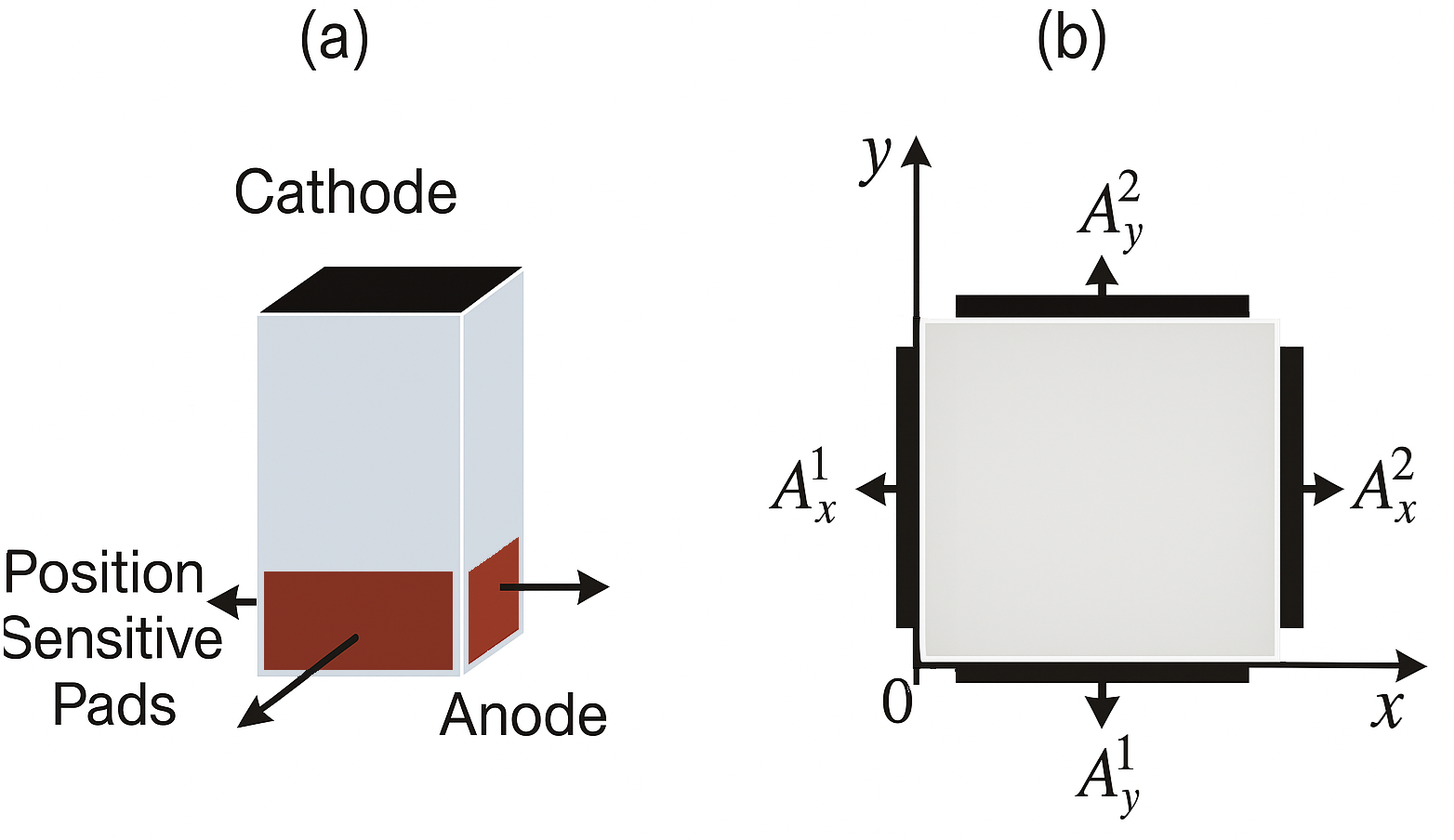}
    \caption{Schematic of a position-sensitive VFG detector (left) and its coordinate system (right)\,\cite{Bolotnikov2020a}.}
    \label{vfg}
\end{figure}

The first generation of VFG detectors had a simple design (see Fig.~\ref{vfg}). The crystals had a high aspect ratio (bars) and were furnished with two monolithic contacts serving as the cathode and anode. The encapsulated bars were enclosed in ultra-thin polyester shells with 5-mm-wide shielding collars placed near the anodes. The detectors were integrated into 6\,×\,6 arrays inside a honeycomb-type holder with spring connectors for easy insertion and replacement of individual detectors. The array module was coupled with a BNL-designed front-end analog AVG1 ASIC. The ASIC has 36 anode and 9 cathode inputs. The 36 detectors in the 6\,×\,6 array were grouped into nine 2\,×\,2 subarrays with common cathodes, whose signals were used to measure Z-coordinates (depths of interactions) and correct charge losses due to electron trapping. This technique is referred to as 1D correction. The energy resolutions achieved with these arrays were better than 1.5\% FWHM at 662\,keV (after 1D corrections).

As the next step in advancing the VFG design, charge-sensing pads were introduced to enable full 3D position sensitivity and allow for correction of response non-uniformities in all three dimensions \cite{Bolotnikov2023}. Position-sensitive VFG detectors employ a shielding electrode comprised of four separate pads, each on one side of the detector \cite{Lutz2007}. The transient signals captured from position-sensitive pads are processed to evaluate the X-Y coordinates of interaction sites, while the measured drift time or the cathode-to-anode ratio (C/A) is used to evaluate the Z-coordinates. For this type of detector, the next generation of ASICs - AVG2 was developed at BNL.

The AVG2 ASIC had the ability to capture and process signals of both polarities since the pad signals could be positive or negative. Unfortunately, these ASICs employed analog signal processing with onboard shaping amplifiers. Today, analog ASICs are being replaced by those implementing waveform digitization.

\section{Detector Design}
\label{sec:design}
\begin{figure}
    \centering
    \includegraphics[width=1\linewidth]{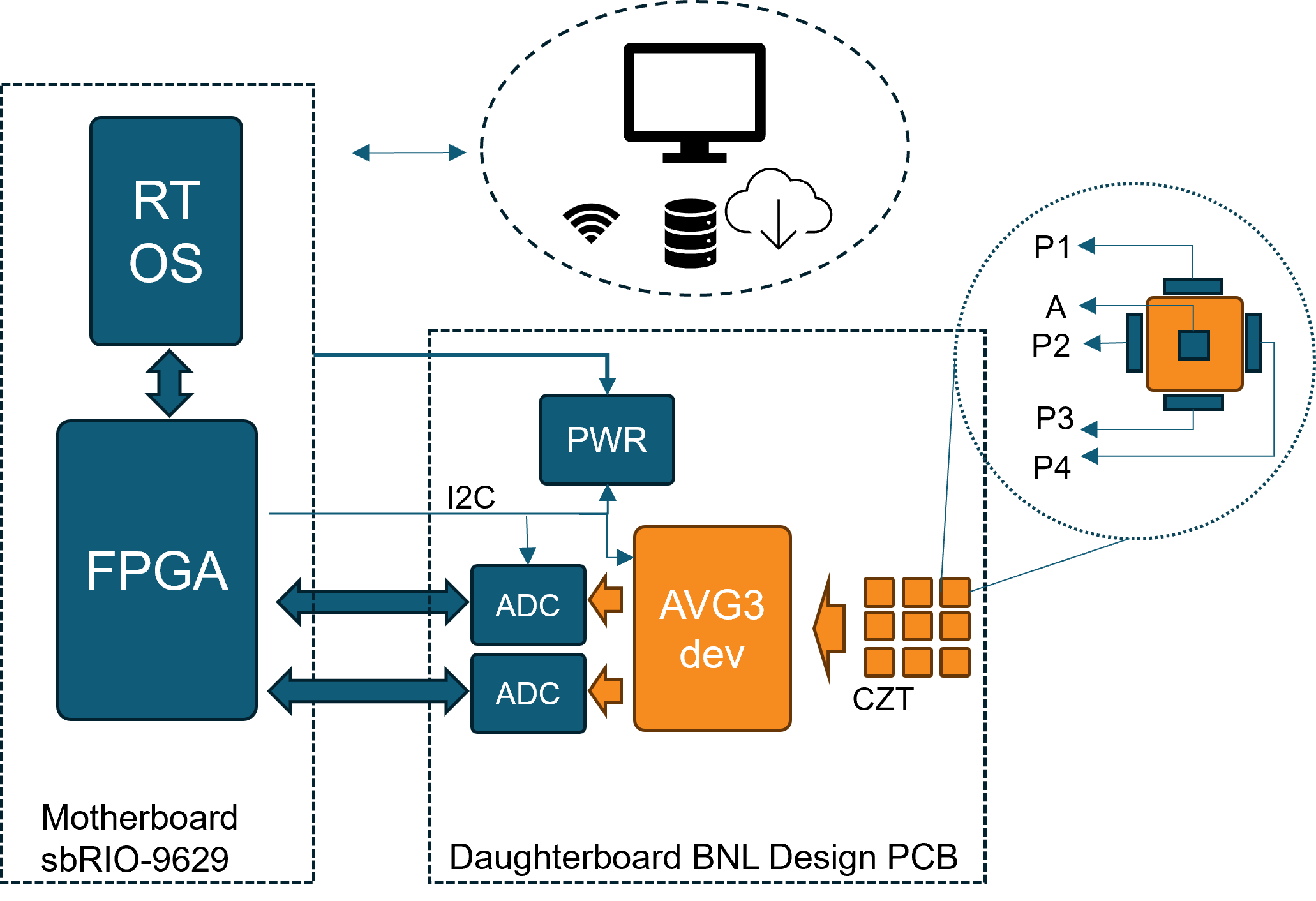}
    \caption{Top-level diagram of the presented detector system.}
    \label{fig:TopSchem}
\end{figure}

The top-level block diagram of the detector is shown in Fig.~\ref{fig:TopSchem}. The system consists of two primary components: a commercial off-the-shelf embedded controller (sbRIO 9629) serving as a motherboard, and a custom, in-house designed daughter PCB containing CZT sensors, multichannel charge-sensitive amplifier (AVG3\_Dev)\,\cite{Pinaroli_2022}, and fast ADCs. The use of a commercial embedded system enables a focused development of the analog front-end and digital signal processing while leveraging a professionally designed and extensively tested digital platform. This approach facilitates full software-based customization, significantly reducing development time during prototyping. Furthermore, if the embedded system meets all operational requirements, it can be seamlessly integrated into the final end-user device. The objective of this work was to design the sensitive analog front-end part of the system and define its functionality through software rather than develop a new embedded platform.  

The modular design, combining a commercial controller with a custom daughter board, simplifies the latter’s design by centering development around the device under test rather than the embedded system itself \cite{Maj_2020}. Consequently, the daughter board consists of only a few essential components:  
\begin{itemize}[itemsep=0pt, topsep=1pt, partopsep=0pt]
    \item A 3\,\(\times\)\,3 array of CZT crystals mounted directly on the PCB;
    \item An AVG3\_Dev ASIC wire-bonded to the PCB, interfacing with the CZT crystals on one end and the ADC on the other;
    \item Two AD9649 ADCs (16-channel, 14-bit, up to 65\,MHz sampling rate) for high-speed digitization of the analog signals from the AVG3\_Dev IC;
    \item Programmable voltage regulators and DAC/ADC for external biasing and measurement;
    \item Debugging pins and indicator LEDs for system diagnostics;
\end{itemize}  

\subsection{Detector front-end: CZT crystals and AVG3\_Dev integrated circuit}
\label{subsec:frontend}

\begin{figure}[!ht]
    \centering
    \includegraphics[width=0.8\linewidth]{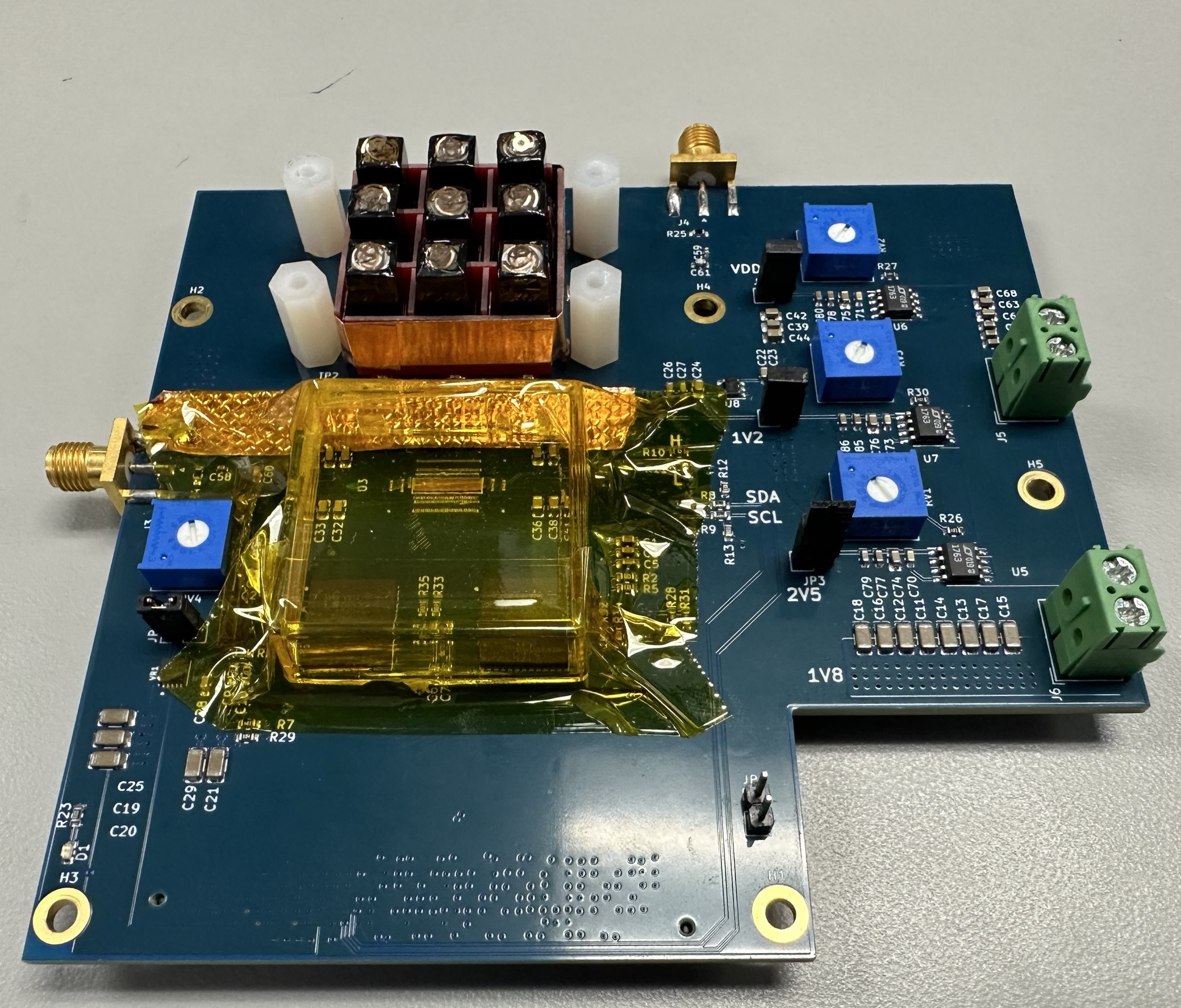}
    \caption{Test hardware set-up consisting of 9 CZT crystals, AVG3\_Dev ASIC and two 16-channel ADCs.}
    \label{fig:CrystalArray}
\end{figure}

The sensitive part of the detector consists of a 3\,\( \times\) \,3 array of 6\,mm\,\( \times\)\,6\,mm\,\( \times\)\, 15\,mm CZT crystals sourced from Redlen Technologies (see Fig.~\ref{fig:CrystalArray}). 
Each crystal is electrically connected to a multichannel charge-sensitive amplifier, AVG3\_Dev, through five signal paths: a central anode (A), which collects the charge generated within the active volume of the crystal, and four side pads (P1-P4), that register the induced signal as the charge drifts toward the anode. To optimize signal routing, side pads are shared between adjacent crystals reducing the total number of connections to 29, while reserving three additional signals for external calibration and debugging.  

\begin{figure}[htbp]
     \centering
     \begin{subfigure}[b]{0.45\textwidth}
         \centering
         \includegraphics[width=0.65\textwidth]{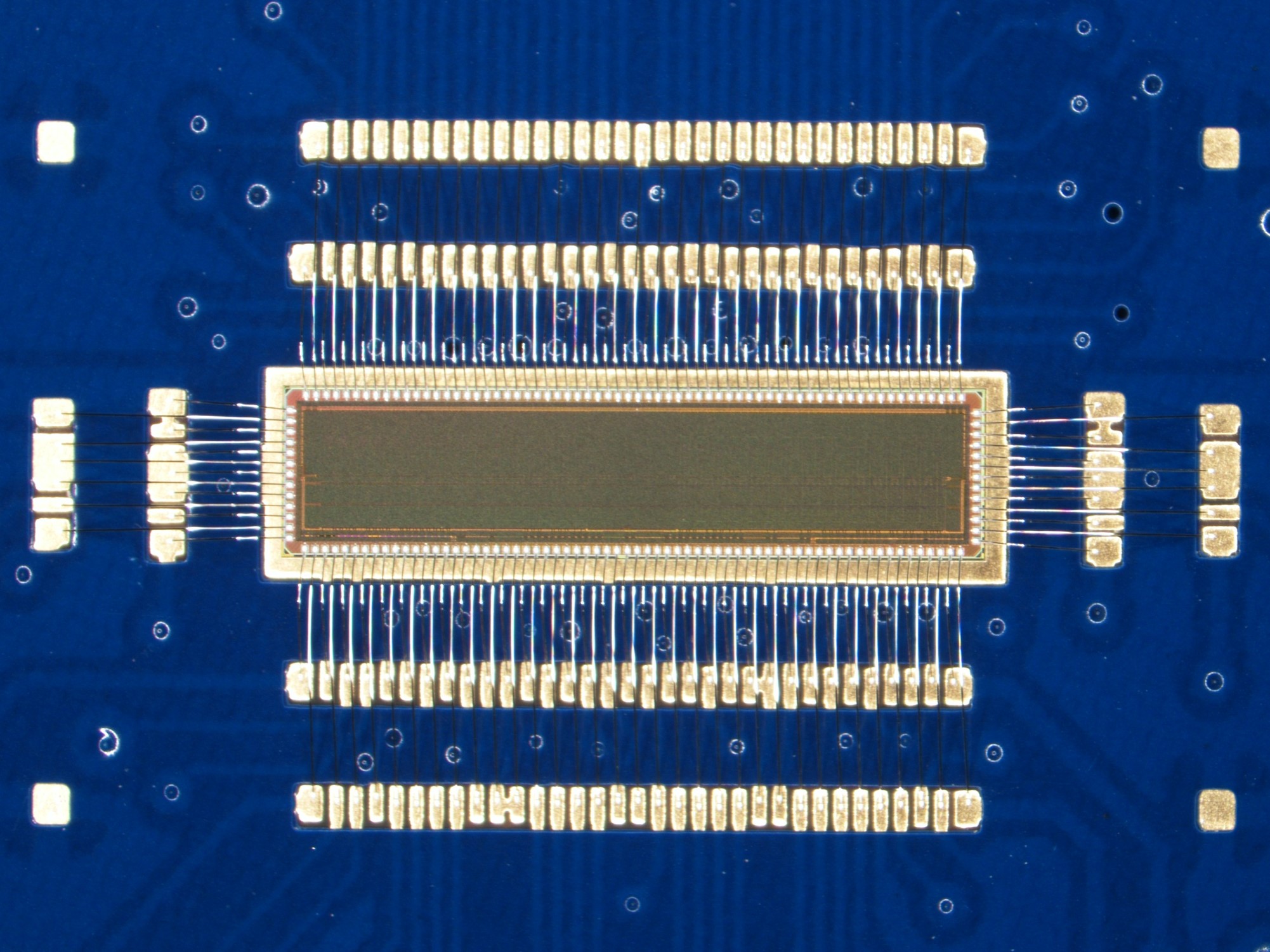}
         \caption{}
         \label{fig:board}
     \end{subfigure}
     %\hfill
     \begin{subfigure}[b]{0.45\textwidth}
         \centering
         \includegraphics[width=0.9\textwidth]{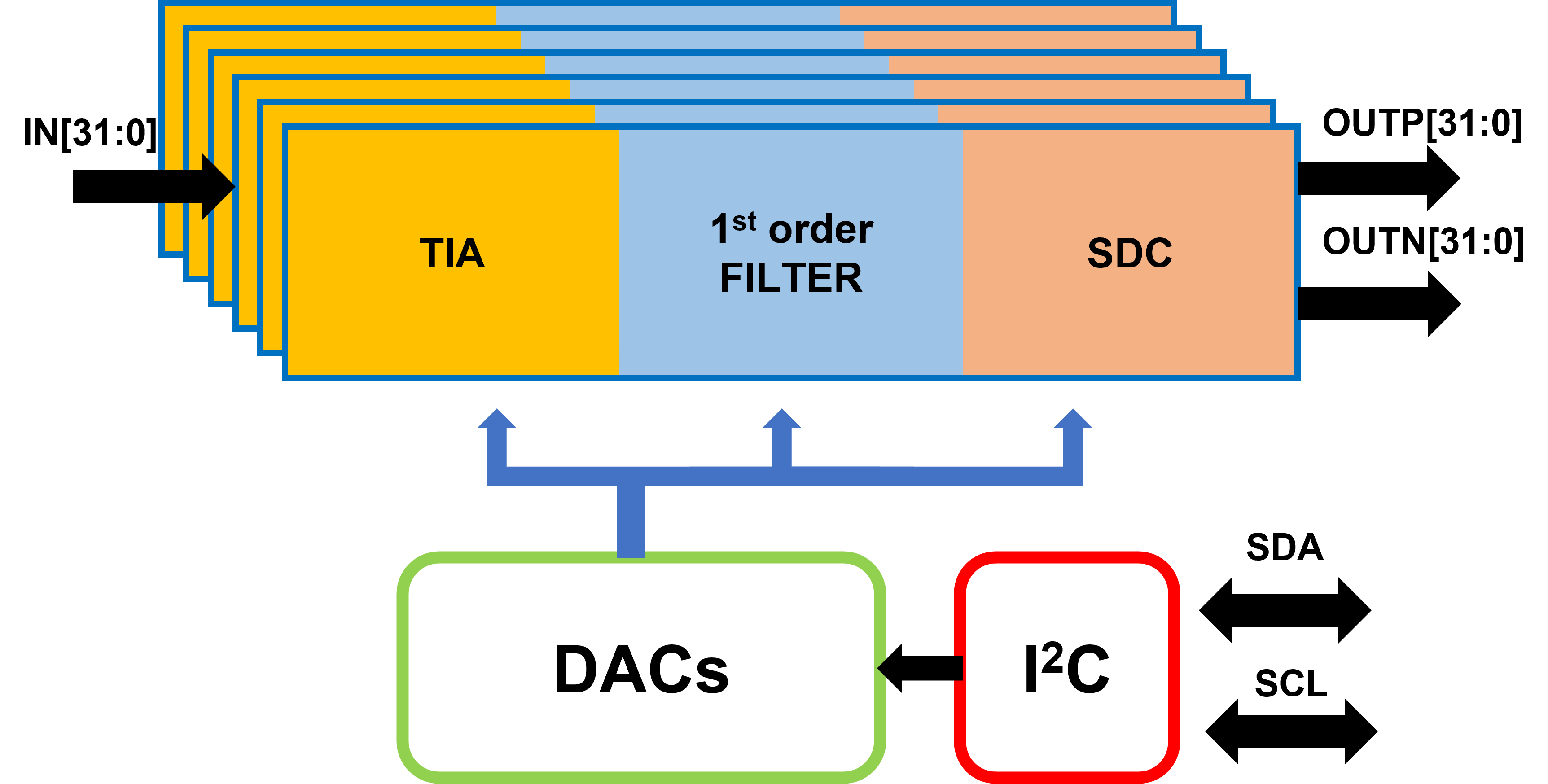}
         \caption{}
         \label{fig:AVG_block}
     \end{subfigure}
        \caption{(a) AVG3\_{}Dev ASIC wire-bonded on the test board, (b) block diagram of the ASIC. Figures from \cite{Pinaroli_2022}.}
        \label{fig:AVG3}
\end{figure}

At the core of the system is the 32-channel semiconductor sensor readout integrated circuit, AVG3\_Dev, designed at Brookhaven National Laboratory\,~\cite{Pinaroli_2022}. It was designed to amplify the charge and shape the signal according to time and noise requirements. Readout channels are optimized for fast pulse shaping from CZT detectors and it is designed to accommodate sensor capacitance in the range of 5\,pF. The circuit can be externally configured via an I2C interface to optimize the signal-to-noise ratio. The bare die chip and the block diagram of the ASIC are shown in Fig.~\ref{fig:AVG3}.  

The AVG3\_Dev outputs a differential analog voltage with a bandwidth of up to 10~MHz, enabling precise pulse shape estimation. To prevent aliasing noise, high-speed data acquisition is required at 40~MHz or higher. The AVG3\_Dev's analog output signals are digitized using high-speed, high-resolution ADCs. Given the high bandwidth of the AVG3\_Dev, precise power management and signal routing are essential. Noise suppression and thermal stability were key design considerations, with all IC connections optimized to minimize parasitic effects.  

\begin{figure}
    \centering
    \includegraphics[width=0.8\linewidth]{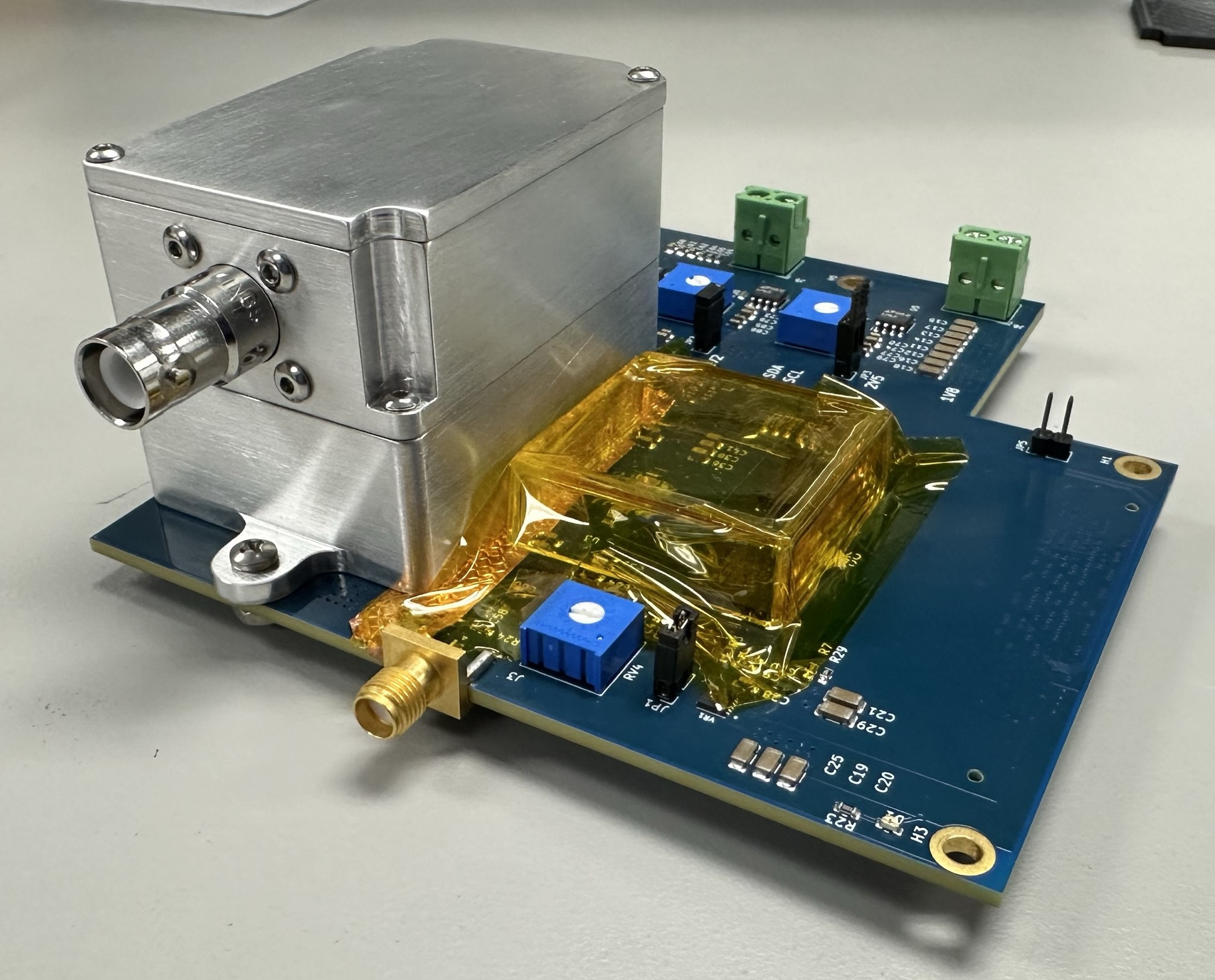}
    \caption{Test hardware set-up. The metal enclosure contains nine CZT crystals and the AVG3\_Dev ASIC is protected by a plastic cover.}
    \label{fig:DetectorEnclosure}
\end{figure}

The daughter PCB is designed with a strong emphasis on noise immunity and signal integrity. Buried wiring between ground layers minimizes noise pickup, while careful PCB routing reduces crosstalk between adjacent channels. High-frequency signals are properly decoupled to mitigate electromagnetic interference (EMI). The CZT crystals are housed in a metal enclosure, providing mechanical protection and shielding against external interference (see Fig.~\ref{fig:DetectorEnclosure}). The enclosure provides a SHV connector for the high voltage detector bias, normally operated at 2\,kV at room temperature.
A key feature of the daughter board architecture is the implementation of a virtual Frisch grid, where anode and pad signals are combined to determine the gamma photon interaction position within the crystals. This method significantly enhances spatial resolution and energy measurement accuracy, allowing for signal directionality determination \cite{Bolotnikov2016}. Data processing is performed in the digital domain using FPGA-based computations on 29 simultaneously sampled signals, grouped into nine crystals, each consisting of five signals (with some signals shared between adjacent groups).

\subsection{Detector Digital Backend}
\label{subsec:backend}

The embedded controller interfaces with the daughter-board via a high-density, 360-pin connector, transmitting power, high-speed and low-speed digital signals. As our design utilizes 32 ADC channels having 14-bit resolution and sampled at 50\,MHz, this configuration yields a combined data stream of 22.4\,Gbps. The data throughput imposes strict requirements on the embedded controller to efficiently handle large data volumes. The sbRIO-9629, equipped with an Artix-7 FPGA and an Intel Atom CPU, meets these demands, offering both the processing power necessary for real-time data management and advanced debugging and monitoring capabilities. By focusing on software-defined functionality, the system achieves flexibility without sacrificing performance. The FPGA is configured to perform the following key operations:
\begin{itemize}[itemsep=0pt, topsep=1pt, partopsep=0pt]
    \item I2C interface state machine with a configurable clock and data bytes;
    \item High-speed interface to the AD9649 ADC, supporting reprogrammable sampling frequencies of up to 65\,MHz, auto-bit-slip and programmable data delay to ensure proper synchronization of all digital lines;
    \item Digital Signal Processing path operating in point-by-point manner, described later in the article;
    \item Buffered Direct Memory Access (DMA) channel data transfer to the RAM memory;   
\end{itemize}

\begin{figure}
    \centering
    \includegraphics[width=1.0\linewidth]{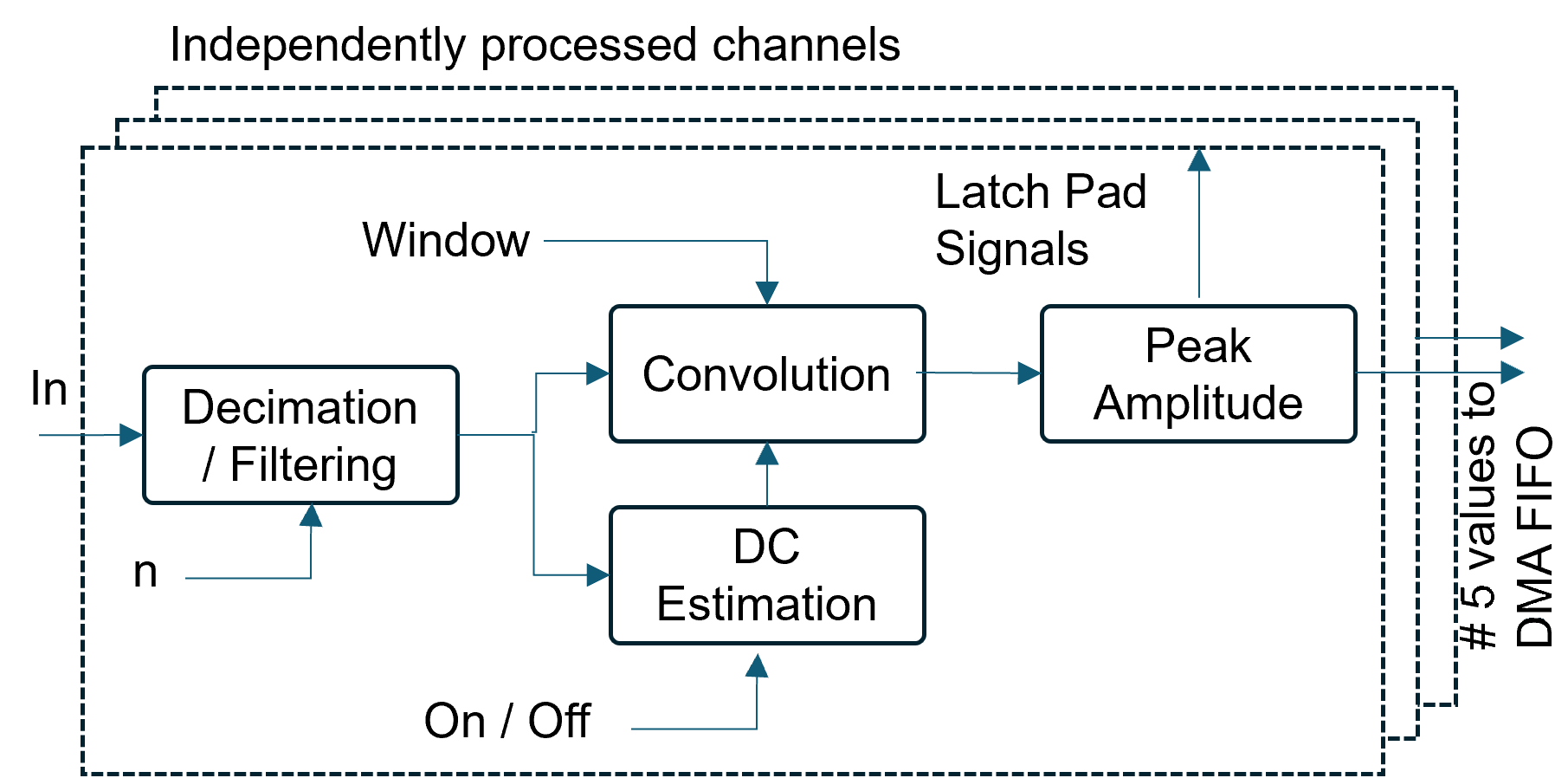}
    \caption{Digital Signal Processing Blocks in FPGA.}
    \label{fig:FPGA-DSP}
\end{figure}

Digital signal processing is implemented entirely within the FPGA for real-time performance. The digital signal processing path is schematically shown in Fig.~\ref{fig:FPGA-DSP}. 
All operations are performed using point-by-point paradigm with a constant-length buffer for convolution. Every new value acquired is put to the buffer while the oldest one is removed. This point-by-point processing avoids bottlenecks associated with waveform snippet transfer and ensures high-speed operation, allowing for dynamic reconfiguration of crucial parameters like signal integration time, convolution window shape, trigger level, DC component estimation and subtraction, etc. Main processing blocks include:
\begin{itemize}[itemsep=0pt, topsep=1pt, partopsep=0pt]
    \item Decimation and Filtering: Enhances signal-to-noise ratio (SNR) by integrating samples over a programmable window, effectively filtering high-frequency noise. The module is configured to keep chosen resolution as it can remove desired number of LSB, ensuring lower FPGA utilization; 
    \item DC Estimation for offset Removal: This block continuously monitors mean value of a signal and its variance to identify and mitigate noise bursts and pulse occurrence prior to amplitude measurement. The block can be switched on or off;
    \item Convolution for processing anode and pad signals using per-channel selectable 32-sample window optimized for their respective signal characteristics;
    \item Anode signal peak/valley detection and adjustable delay for pad-signal value latching, allowing optimization of reconstruction of the point of interaction;
    \item Buffered DMA FIFO transfer of five values: an estimated anode peak amplitude and four corresponding pad-signals values;
    \item Continuous raw data and triggered data snippets streaming from a dynamically selected single channel;
\end{itemize}

The FPGA firmware enables dynamic parameter adjustment, allowing real-time optimization for specific measurement scenarios.  
The detector operation and data storage are managed by an application running on a customized Linux Real-Time operating system. This application consists of multiple tasks, with the most critical ones responsible for controlling FPGA operation (e.g., Start, Stop, Status) and receiving an in-FPGA computed values via a DMA channel.  

Each recorded event comprises five values: the anode amplitude and four corresponding pad values, enabling precise 2D event localization. This approach enables the effective segmentation of a single crystal into sub-crystals. The presented detector supports a single crystal grid division up to 32\,\(\times\)\,32, assigning each event to one of 1,024 sub-pixel positions. For each position, a sub-pixel energy spectrum is recorded, consisting of 3,000 points, each represented as a 4-byte value. As a result, the program allocates over 110\,MB of RAM upon initialization for temporary spectral storage, while preserving the ability to log every event with a corresponding timestamp.  

The application also provides an option to stream raw or processed data from a selected channel for debugging purposes. Additionally, it features an embedded graphical user interface (GUI) for real-time data visualization and interactive control. Finally, the system supports both wired and wireless connections to a host PC or network storage, enabling remote data access and management.  

%\begin{figure*}
%    \centering
%    \includegraphics[width=0.7\linewidth]{Picts/ApplicationScreen.png}
%    \caption{Main Application GUI with \isotope[137]{Cs} and \isotope[133]{Ba} raw data spikes and %a combined spectrum}
%    \label{fig:GUI}
%\end{figure*}

\section {System validation with gamma sources}
\label{sec:verification}

The detector was prepared for testing and validation. A dedicated application was deployed enabling the configuration of the ADC and AVG3\_Dev parameters, as well as adjustments of digital signal processing settings. The application features an embedded graphical user interface (GUI)
% showed in in Fig.~\ref{fig:GUI} 
and data storage functionality. The CZT crystals were biased at -2,000\,V, and both raw data streams and selected data snippets were recorded for detailed analysis. 

\subsection{Noise performance estimation}
\label{subsec:noise}

The detector was tested to ensure proper data flow and to evaluate the accuracy of its digital signal processing, for noise performance, throughput and uniformity with all 9 crystals operating simultaneously, as well as for high-energy gamma photon detection (around the 2\,MeV range). To perform those tests different sealed gamma sources of \isotope[137]{Cs}, \isotope[133]{Ba} and \isotope[228]{Th} were placed on top of the metal enclosure housing the crystals. 

\begin{figure}[htbp]
     \centering
     \begin{subfigure}[b]{0.5\textwidth}
         \centering
         \includegraphics[trim={0.55cm 0.4cm 0.5cm 0.5cm},clip,width=0.85\textwidth]{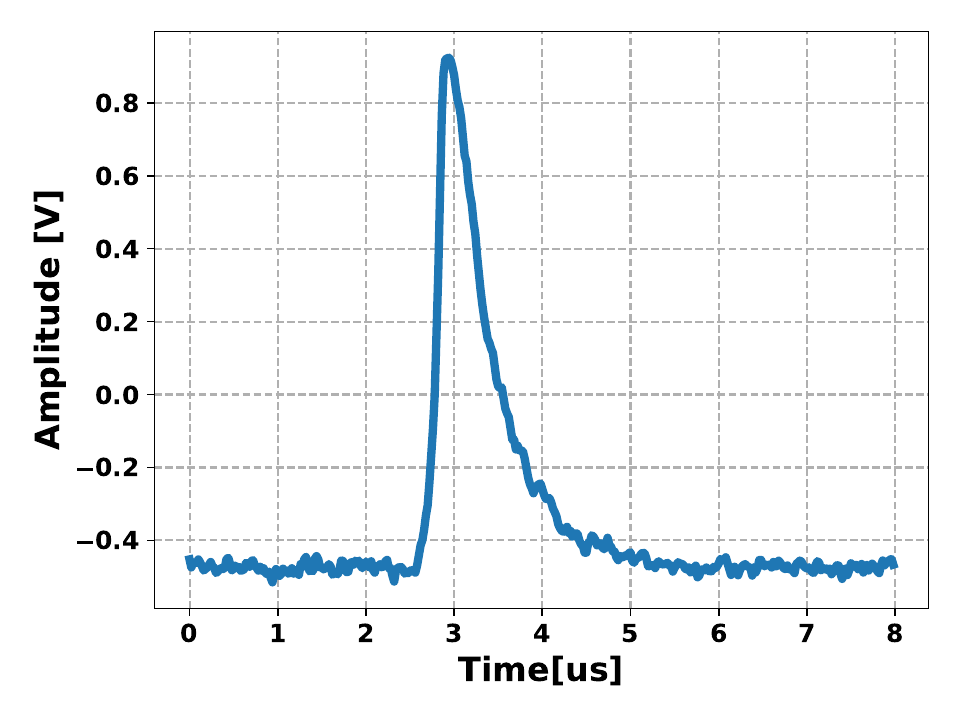}
         \caption{}
         \label{fig:raw}
     \end{subfigure}
     %\hfill
     \begin{subfigure}[b]{0.5\textwidth}
         \centering
         \includegraphics[trim={0.55cm 0.6cm 0.5cm 0.5cm},clip,width=0.85\textwidth]{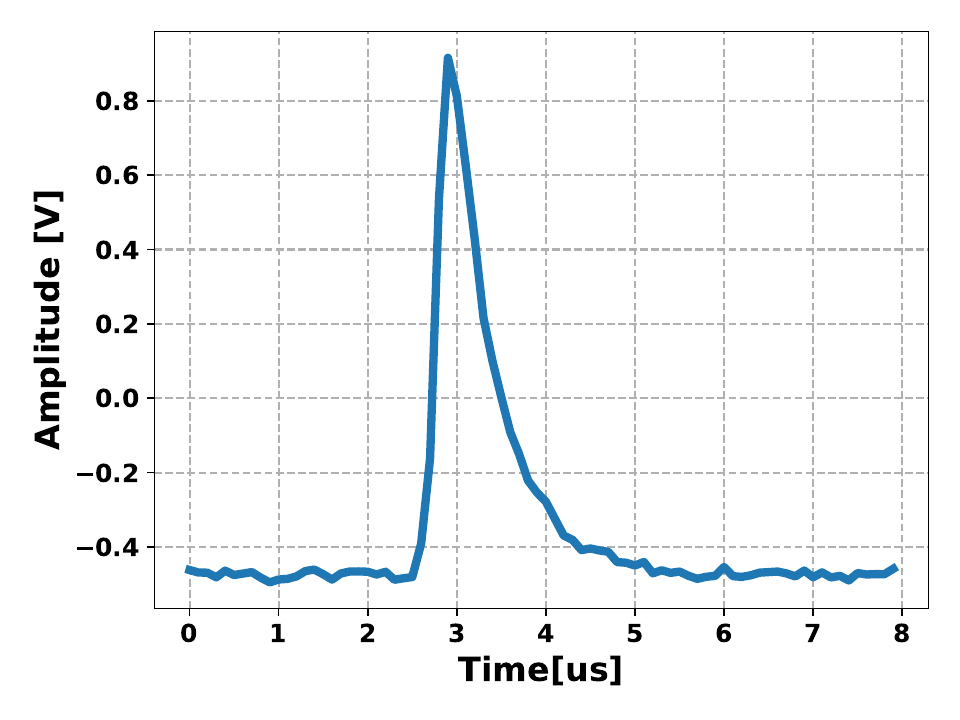}
         \caption{}
         \label{fig:filter}
     \end{subfigure}
     %\bigskip
     \begin{subfigure}[b]{0.5\textwidth}
          \centering
         \includegraphics[trim={0.4cm 0.4cm 0.35cm 0.5cm},clip,width=0.85\textwidth]{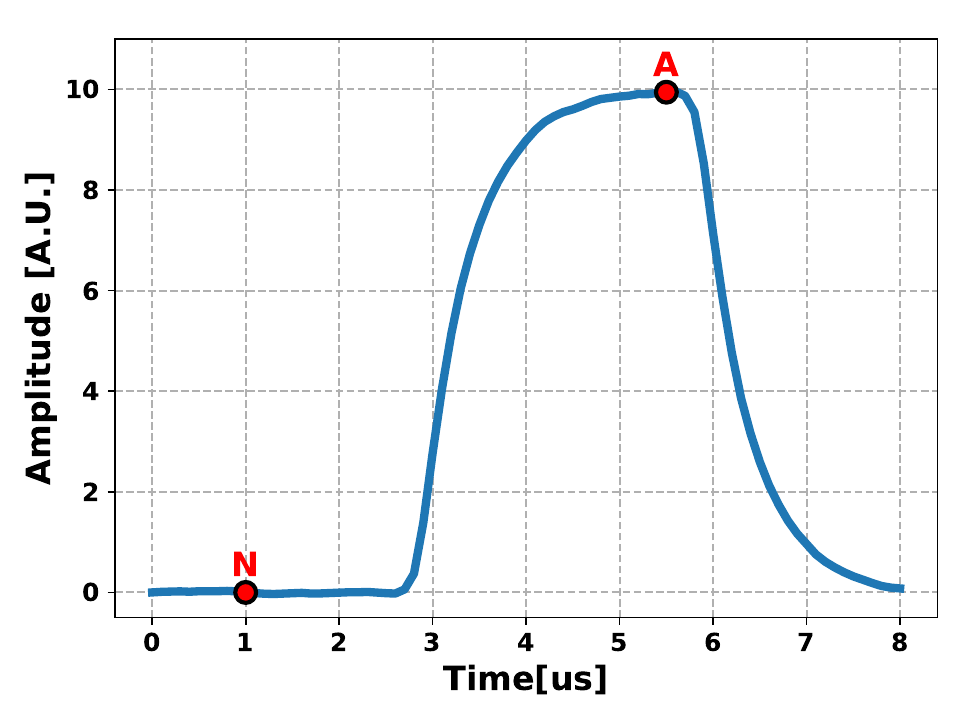}
         \caption{}
         \label{fig:integral}
     \end{subfigure}
        \caption{Illustration of the signal processing steps: (a) Raw signal, (b) Signal after decimation, (c) Final processed signal after convolution and DC offset removal.}
        \label{fig:signal_processing}
\end{figure}

Exemplary waveforms registered along the digital signal processing data flow are shown in Fig.~\ref{fig:signal_processing}. The raw waveform acquired by an anode of a single crystal without any corrections is presented in Fig.~\ref{fig:signal_processing}a, while Fig.~\ref{fig:signal_processing}b illustrates the same signal after in-FPGA decimation block, effectively reducing high-frequency noise. Finally, Fig.~\ref{fig:signal_processing}c displays the result after convolution with rectangular window. At this stage of signal processing, the pulse shape is transformed, and the amplitude represents the detected energy using arbitrary units, which must be scaled accordingly, depending on the DSP parameter settings. 

%\FloatBarrier  % Forces figures to stay within this subsection

%\iffalse

%\fi

\begin{figure}[!ht]
    \centering
    \includegraphics[trim={0.55cm 0.6cm 0.1cm 0.3cm},clip,width=0.9\linewidth]{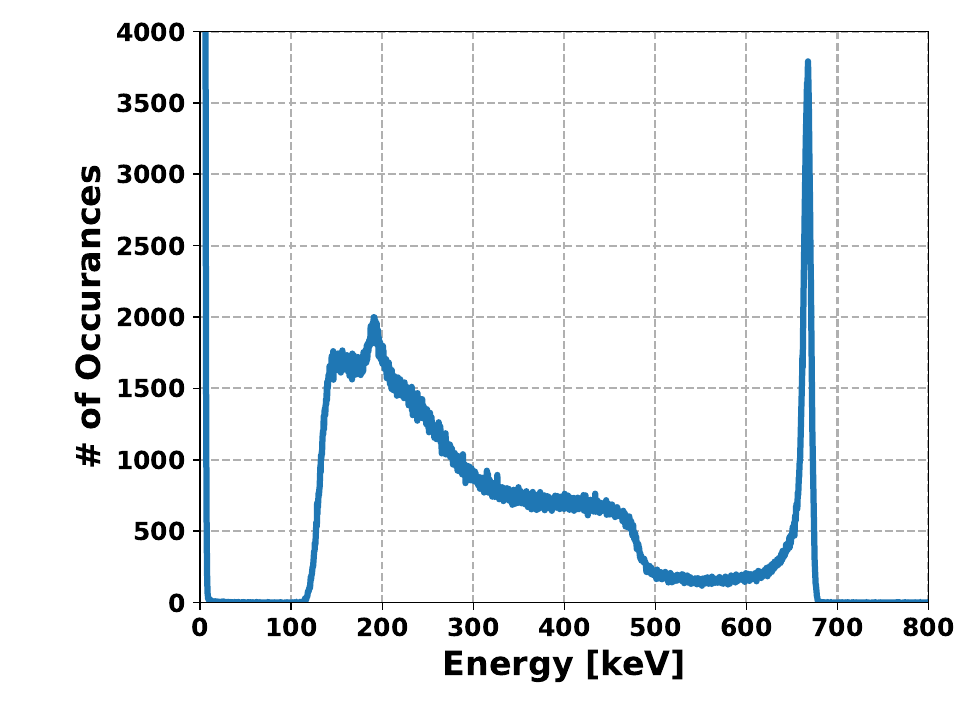}
    \caption{\isotope[137]{Cs} spectrum collected by a single crystal. The most left peak represents the noise counts.}
    \label{fig:cs137}
\end{figure}

The signal peak amplitude (A) is identified in the FPGA and acquired for radioisotope energy spectrum calculation. For the purpose of electronic channel noise estimation, a corresponding noise value (N) is also acquired together with the triggered peak amplitude (Fig.~\ref{fig:cs137}). This way, we are able to differentiate the energy resolution component related to the electronic noise of the designed channel and the noise originating from crystal imperfections and in-crystal charge transfer.

The energy resolution of the primary peak was estimated by fitting its right-side profile to a Gaussian function, as the left side exhibits significant background. The calculated resolution of the spectrum shown in Fig.~\ref{fig:cs137} is approximately 1.18\%, while the corresponding channel noise level measured at the same time is only 0.59\%, which is negligible taking into account that noise components are added in a square root. These results suggest that system performance concerning the input signal having a 6\,mm\,\(\times\)\,6\,mm\,\(\times\)\,15\,mm CZT crystal attached and properly biased is capable of measuring signals with as good as 0.6\% resolution. Using this methodology we can assess that the noise is dominated by the crystal imperfections \cite{Bishop2012, Kim2015}, and therefore, it can further be improved without any changes to the system design, but by better selection of crystals or by applying the so-called 2D and 3D correction to the spectrum. The correction is made by estimation of an event position in 2D (pixel) or in 3D (voxel) and then projection of a spectrum for a single sub-voxel. The correction methods are described in detail in \cite{Bolotnikov2020b}. 

\subsection{Uniformity and throughput estimation}
\label{subsec:uniformity}
\begin{figure}[!ht]
    \centering
    \includegraphics[trim={0.2cm 0.35cm 0.2cm 0},clip,width=0.9\linewidth]{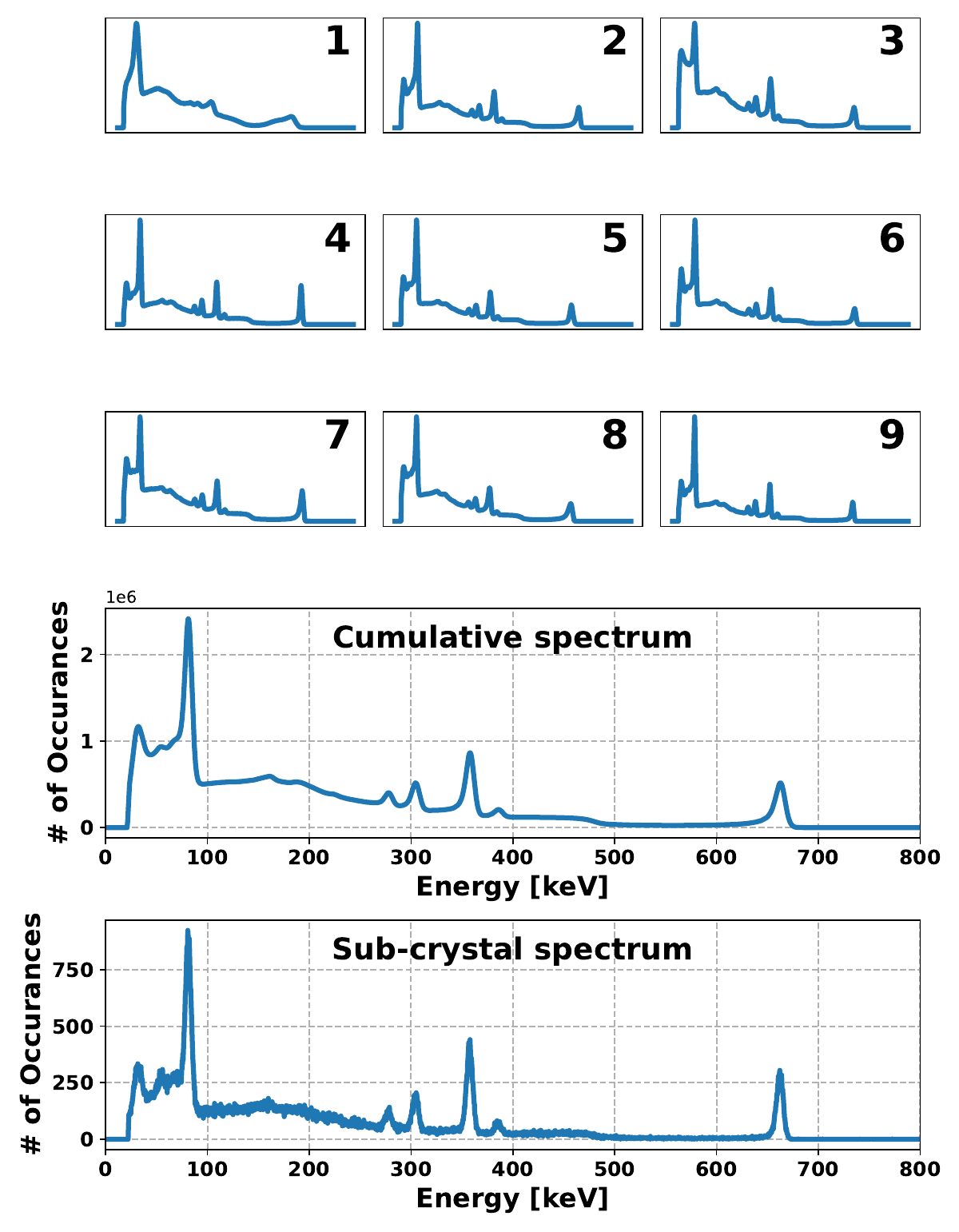}
    \caption{Combined spectrum of \isotope[137]{Cs} and \isotope[133]{Ba} collected by the nine crystals in the assembly. Crystal \#\,1 (top-left corner) shows the poorest response among the used crystals. Bottom plot shows the cumulative spectra from the 8 well responding crystals. The last one is a chosen sub-pixel of a crystal no.\,4 showing 1\% energy resolution.}
    \label{fig:9crystals}
\end{figure}

Assessing the throughput and uniformity of all nine crystals simultaneously is crucial for the implementation of source localization and overall detection efficiency. Uniform data across all channels is essential for summing the signals and obtaining a combined spectrum, a factor that becomes even more significant for larger detector arrays. To evaluate uniformity, two gamma isotopes \isotope[137]{Cs} and \isotope[133]{Ba}, having more energy emission lines, were placed on top of the metal shield. The individual spectra are presented in Fig.~\ref{fig:9crystals}.
 
\begin{comment}

\begin{figure}
    \centering
    \includegraphics[trim={0.2cm 0.35cm 0.2cm 0},clip,width=0.9\linewidth]{Picts/9crystal_uncorrectedA.pdf}
    \caption{\isotope[137]{Cs} spectrum collected by the nine crystal in the assembly. Crystal \#\,1 (top-left corner) shows the poorest response among the used crystals.}
    \label{fig:9crystals}
\end{figure}

\begin{figure}[!h]
    \centering
    \includegraphics[width=0.9\linewidth]{Picts/spectra_9crystal.pdf}
    \caption{\isotope[137]{Cs} summed spectrum collected by eight well-performing crystals in the assembly.}
    \label{fig:9crystalsSum}
\end{figure}
\end{comment}
Due to its noticeably poor performance, crystal no.~1 was excluded from the uniformity analysis. After trimming all 8 remaining crystal's data, we integrate them together to obtain a clean spectrum shown in Fig.\,\ref{fig:9crystals}, where primary energies of both isotopes are clearly visible and a very narrow noise part (estimated with (N) points from  Fig.~\ref{fig:signal_processing}\,(c)) is visible.

Within this measurement, the anode signal amplitude is recorded together with the four corresponding pad signals, enabling two-dimensional event localization. Having the 3\,\(\times\)\,3 crystal matrix and the 32\,\(\times\)\,32 virtual subdivision per crystal, the system generates over 9,000 individual spectra, each containing 3,000 data points. Due to the low activity of the non-collimated sealed sources, acquiring statistically significant spectra for each sub-pixel requires tens of hours of data collection. The exemplary sub-pixel spectrum demonstrating sub-1\% resolution is illustrated in Fig.\,\ref{fig:9crystals} (bottom). This allows to further increase the noise performance and the corresponding energy resolution.

\subsection{Performance with 2\,MeV gamma photons}
\label{subsec:performance}
\begin{figure}[!ht]
    \centering
    \includegraphics[trim={0.7cm 0.35cm 0.3cm 0.20cm},clip,width=1.0\linewidth]{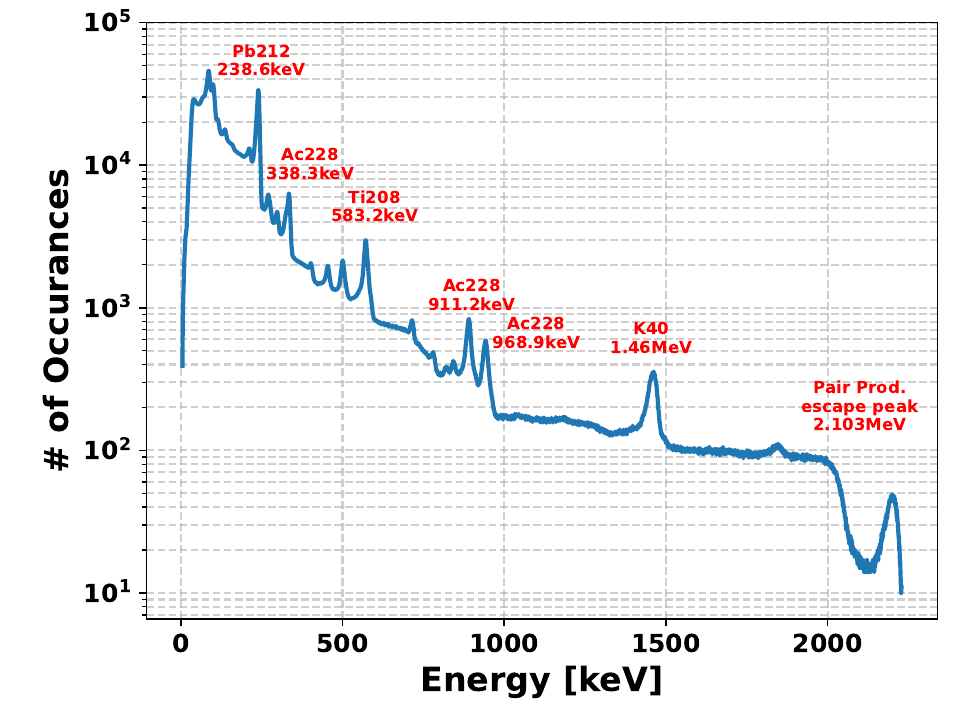}
    \caption{Cumulative spectrum from 8 crystals irradiated with an uncollimated \isotope[228]{Th} radioactive source.}
    \label{fig:Thorium}
\end{figure}

The performance of the detector capturing gamma photons in the 2\,MeV range was tested with \isotope[228]{Th} sealed source placed nearby. The integrated spectrum is shown in Fig.~\ref{fig:Thorium}. Clear visibility of details in the spectrum demonstrates low-noise and high dynamic range performance. The range is limited by the DSP algorithm, namely the length of a convolution window. Extending the dynamic range is possible by configuring longer window or changing the decimation parameters effectively shortening the pulse before convolution operation, which would sacrifice the noise performance, but would extend the dynamic range.

\section{Conclusions}
\label{sec:conclusion}

This work presents the design, development, and performance evaluation of a Virtual Frisch-Grid geometry CZT gamma-ray detector, demonstrating its capability for precise radioisotope identification across a broad energy range. The detector, built with a 3\,\(\times\)\,3 matrix of CZT crystals and integrated with a high-performance AVG3\_Dev readout IC, leverages FPGA-based digital signal processing to enable real-time operation, high-speed data acquisition, and event localization. Our approach combines a commercial embedded platform with a custom analog front-end, allowing efficient prototyping while maintaining flexibility for future system refinements.

Performance testing confirmed that the system achieves an energy resolution of approximately 1.18\%, with electronic noise contributions as low as 0.59\%, highlighting the potential for further optimization through crystal selection and advanced spectral corrections. This low noise performance is achieved through in-FPGA DSP algorithms, which effectively suppress noise while maintaining high signal fidelity. Additionally, multi-crystal measurements and sub-pixel event reconstruction demonstrated the capability of the system for high-throughput analysis and improved spatial resolution, yielding over 9,000 individual spectra. The successful operation of the detector system with high-energy gamma sources, such as \isotope[228]{Th}, further validates its dynamic range and low-noise performance. These results suggest that our detector provides a viable, field-deployable solution for radioisotope identification, with potential applications in nuclear safety, environmental monitoring, and material characterization. Future work will focus on refining spectral correction techniques and extending the system’s dynamic range through enhanced DSP configurations, ultimately advancing its capabilities for precision gamma-ray spectroscopy.

% Remaining Citations ~\cite{Voyce2021}, ~\cite{Zhang2020},

\bibliographystyle{elsarticle-num} 
\bibliography{biblio.bib}

\begin{thebibliography}{10}
\expandafter\ifx\csname url\endcsname\relax
  \def\url#1{\texttt{#1}}\fi
\expandafter\ifx\csname urlprefix\endcsname\relax\def\urlprefix{URL }\fi
\expandafter\ifx\csname href\endcsname\relax
  \def\href#1#2{#2} \def\path#1{#1}\fi

\bibitem{97e69526cbca4e83bb52a5e7ebe8f16d}
M.~Nogami, K.~Hitomi, T.~Onodera, K.~Watanabe, K.~Ishii, Reversible capacitive
  frisch grid tlbr detectors, Japanese journal of applied physics 62~(9),
  publisher Copyright: {\textcopyright} 2023 The Japan Society of Applied
  Physics. (Sep. 2023).
\newblock \href {https://doi.org/10.35848/1347-4065/acf0aa}
  {\path{doi:10.35848/1347-4065/acf0aa}}.

\bibitem{HITOMI1999160}
K.~Hitomi, O.~Muroi, T.~Shoji, T.~Suehiro, Y.~Hiratate,
  \href{https://www.sciencedirect.com/science/article/pii/S0168900299006142}{Room
  temperature x and \textgamma-ray detectors using thallium bromide crystals},
  Nuclear Instruments and Methods in Physics Research Section A: Accelerators,
  Spectrometers, Detectors and Associated Equipment 436~(1) (1999) 160--164.
\newblock \href {https://doi.org/https://doi.org/10.1016/S0168-9002(99)00614-2}
  {\path{doi:https://doi.org/10.1016/S0168-9002(99)00614-2}}.
\newline\urlprefix\url{https://www.sciencedirect.com/science/article/pii/S0168900299006142}

\bibitem{Bolotnikov2020b}
A.~E. Bolotnikov, J.~MacKenzie, E.~Chen, F.~J. Kumar, S.~Taherion, G.~Carini,
  G.~D. Geronimo, J.~Fried, K.~Kim, L.~O. Girado, E.~Vernon, R.~B. James,
  Performance of 8×8×32 and 10×10×32 mm3 cdznte position-sensitive virtual
  frisch-grid detectors for high-energy gamma-ray cameras, Nuclear Instruments
  and Methods in Physics Research A 969 (2020) 164005.

\bibitem{Lordi2013}
V.~Lordi, Point defects in cd(zn)te and tlbr: Theory, Journal of Crystal Growth
  379 (2013) 84--92.

\bibitem{H3Dref}
W.~Kaye, D.~Barron, F.~Zhang, M.~Streicher, H.~Yang, A.~Alawi, K.~Moran, Z.~He,
  Quantitative analysis using a compact high resolution gamma-ray spectrometer,
  in: 2020 IEEE Nuclear Science Symposium and Medical Imaging Conference
  (NSS/MIC), 2020, pp. 1--5.
\newblock \href {https://doi.org/10.1109/NSS/MIC42677.2020.9508030}
  {\path{doi:10.1109/NSS/MIC42677.2020.9508030}}.

\bibitem{Bolotnikov2023}
A.~E. Bolotnikov, C.~A. Brown, G.~A. Carini, J.~Christian, L.~Cirignano, C.~R.
  Deane, A.~Dellapenna, et~al., Using 3d position sensitivity to reveal
  response non-uniformities in cdznte, tlbr, and cspbbr3 detectors, Nuclear
  Instruments and Methods in Physics Research Section A: Accelerators,
  Spectrometers, Detectors, and Associated Equipment 1057 (2023) 168785.

\bibitem{Lutz2007}
G.~Lutz, Semiconductor Radiation Detectors: Device Physics, Springer, Berlin,
  2007, no. PUBDB-2020-02521.

\bibitem{Bolotnikov2020a}
A.~E. Bolotnikov, G.~S. Camarda, G.~D. Geronimo, J.~Fried, R.~B. James, A 4×4
  array module of position-sensitive virtual frisch-grid cdznte detectors for
  gamma-ray imaging spectrometers, Nuclear Instruments and Methods in Physics
  Research A 954 (2020) 161036.

\bibitem{Pinaroli_2022}
G.~Pinaroli, S.~Herrmann, S.~Miryala, V.~Manthena, G.~Deptuch, G.~Carini,
  A.~Bolotnikov, A.~Dellapenna, E.~Raguzin, J.~Fried, C.~Deane, C.~Brown,
  J.~Christian, L.~Cirignano, A.~Kargar, H.~Kim, K.~Shah, M.~Squillante,
  M.~Squillante, E.~Weststrate, A.~Valente, M.~Koslowsky, A.~Miller, M.~Smith,
  \href{https://dx.doi.org/10.1088/1748-0221/17/02/C02011}{Multi-channel
  front-end asic for a 3d position-sensitive detector}, Journal of
  Instrumentation 17~(02) (2022) C02011.
\newblock \href {https://doi.org/10.1088/1748-0221/17/02/C02011}
  {\path{doi:10.1088/1748-0221/17/02/C02011}}.
\newline\urlprefix\url{https://dx.doi.org/10.1088/1748-0221/17/02/C02011}

\bibitem{Maj_2020}
P.~Maj, P.~Otfinowski, A.~Koziol, D.~Gorni, Q.~Zhang, P.~Dudek,
  \href{https://dx.doi.org/10.1088/1748-0221/15/03/C03010}{1.2 mfps standalone
  x-ray detector for time-resolved experiments}, Journal of Instrumentation
  15~(03) (2020) C03010.
\newblock \href {https://doi.org/10.1088/1748-0221/15/03/C03010}
  {\path{doi:10.1088/1748-0221/15/03/C03010}}.
\newline\urlprefix\url{https://dx.doi.org/10.1088/1748-0221/15/03/C03010}

\bibitem{Bolotnikov2016}
A.~E. Bolotnikov, G.~S. Camarda, E.~Chen, R.~Gul, V.~Dedic, G.~D. Geronimo,
  J.~Fried, A.~Hossain, J.~M. MacKenzie, L.~Ocampo, P.~Sellin, S.~Taherion,
  E.~Vernon, G.~Yang, U.~El-Hanany, R.~B. James, Use of the drift-time method
  to measure the electron lifetime in long-drift-length cdznte detectors,
  Journal of Applied Physics 120 (2016) 104507.

\bibitem{Bishop2012}
S.~R. Bishop, H.~L. Tuller, G.~Ciampi, W.~Higgins, J.~Engel, A.~Churilov, K.~S.
  Shah, The defect and transport properties of acceptor doped tlbr: role of
  dopant exsolution and association, Physical Chemistry Chemical Physics
  14~(29) (2012) 10160--10167.

\bibitem{Kim2015}
K.~Kim, S.~Kim, J.~Hong, J.~Lee, T.~Hong, A.~E. Bolotnikov, G.~S. Camarda,
  R.~B. James, Purification of cdznte by electromigration, Journal of Applied
  Physics 117~(14) (2015).

\end{thebibliography}

\end{document}